\documentclass[12pt,preprint]{aastex}

\shorttitle{{\it Spitzer} Starforming Core Near NGC 2264 IRS-2}
\shortauthors{E.T. Young et al.}

\begin{document}


\title{{\it SPITZER} AND MAGELLAN OBSERVATIONS OF NGC 2264:\\ A
REMARKABLE STAR FORMING CORE NEAR IRS-2}


\author{E.T. Young\altaffilmark{1},
P.S. Teixeira\altaffilmark{2,3}, C.J. Lada\altaffilmark{2}, J.
Muzerolle\altaffilmark{1}, S.E. Persson\altaffilmark{4},\\ D.C.
Murphy\altaffilmark{4}, N. Siegler\altaffilmark{1}, M.
Marengo\altaffilmark{2}, O. Krause\altaffilmark{1}, \& A.K.
Mainzer\altaffilmark{5} } \altaffiltext{1}{Steward Observatory,
University of Arizona, 933 N Cherry Ave, Tucson, AZ 85721}
\altaffiltext{2}{Harvard-Smithsonian Center for Astrophysics, Mail
Stop 42, 60 Garden Street, Cambridge, MA 02138}
\altaffiltext{3}{Dept. de F\'{\i}sica, Faculdade de Ci\^encias de
Universidade de Lisboa, Lisbon, Portugal}
\altaffiltext{4}{Observatories of the Carnegie Institution of
Washington, 813 Santa Barbara St, Pasadena, CA 91101}
\altaffiltext{5}{Jet Propulsion Laboratory, Caltech, 4800 Oak Grove
Dr., Pasadena CA 91109}



\begin{abstract}
We analyze {\it Spitzer} and Magellan observations of a star forming
core near IRS-2 in the young cluster NGC 2264.  The submillimeter
source IRAS 12 S1, previously believed to be an intermediate mass
Class 0 object is shown to be a dense collection of embedded, low
mass stars. We argue that this group of stars represents the
fragmenting collapse of a dense, turbulent core, based on a number
of indicators of extreme youth.  With reasonable estimates for the
velocity dispersion in the group, we estimate a dynamical lifetime
of only a few x 10$^{4}$ years. Spectral energy distributions of
stars in the core are consistent with Class I or Class 0
assignments.  We present observations of an extensive system of
molecular hydrogen emission knots.  The luminosity of the objects in
the core region are consistent with roughly solar mass protostars.

\end{abstract}




%
\section{Introduction}
Over the past 20 years, a relatively consistent picture of low-mass
star formation has emerged.  In this scenario, a dense molecular
core becomes gravitationally unstable, collapses, and begins the
process of star formation.  We know, however, that real star
formation is significantly more complex.  In particular, stars do
not form in isolation, and interactions between different protostars
may be quite important in their development.  Recent numerical
simulations by \citet{Bate03}, for example, show that star formation
in a turbulent core is a highly dynamic, interactive process.
Because this early formation activity occurs in regions of very high
obscuration, detailed investigations have been difficult given the
limited sensitivity and angular resolution of available far infrared
capabilities.  The observational situation has dramatically improved
with the launch of {\it Spitzer}.

NGC 2264 has long been one of the touchstones in star formation
studies.  Beginning with the work of \citet{Herbig54} and
\citet{Walker56}, studies at
all available wavelengths have been conducted in this extremely
young cluster.  At a distance of 800 pc, NGC 2264 is close enough
for detailed investigations of the stellar population down to very low
masses.  While more distant than some of the other well studied
star formation regions like $\rho$ Ophiuchi or the Orion molecular
cloud, NGC 2264 displays some of the best examples of the interaction
of the interstellar medium with the young stars.

Investigators at optical wavelengths \citep*{Sung97, Flaccomio99,
Park00, Rebull02, Dahm05} have fit their data to theoretical
isochrones and find typical ages in the 1-3 Myr range for the young
stellar population.  They also find, however, that there is a wide
range in derived ages for the individual stars, going from 10$^{5}$
to nearly 10$^{7}$ years. The region contains numerous IRAS sources,
many of which have been classified as Class I protostars
\citep{MLY89}. Moreover, the presence of Herbig-Haro objects,
molecular outflows \citep{MLS88, Wolf-Chase95}, and Class-I objects
\citep{Wolf-Chase03} attest to continuing, active star formation.

One of these Class I sources, IRS-2 (listed as IRAS 06382+0939 in
the IRAS Point Source Catalog) is located 17\arcmin \/ south of the
O7 star S Mon and 7\arcmin \/ north of the Cone Nebula in a dense
clump of molecular gas. It was identified as a source of
far-infrared radiation by \citet{Sargent84} and by
\citet{Schwartz85}. Using the Kuiper Airborne Observatory,
\citet{Cohen85} found a second peak of far infrared radiation
2\arcmin \/ to the southwest. \citet{Castelaz88} mapped the region
at near infrared wavelengths and associated IRS-2 with a pair of Red
Nebulous Objects (RNO-E and RNO-W).  IRS-2 was designated as IRAS 12
by \citet{MLY89} in their IRAS survey of the NGC 2264 region, and
they classified it as a likely Class I protostar.  This
classification was supported by the association of IRAS 12 with a
high velocity molecular outflow (NGC 2264 Flow D). More recently,
this region has been mapped at higher angular resolution at
submillimeter wavelengths by \citet{Williams02} and
\citet{Wolf-Chase03}.  Those maps show that the IRAS source itself
is weak at 450 and 850 $\mu$m, but that it is part of a complex
collection of submillimeter cores.  The latter authors argue that
these cores have a range of evolutionary stages.

\citet{Teixeira06} presented {\it Spitzer} observations of NGC 2264
IRS-2 region. They found that bright 24 $\mu$m sources were aligned
in a spoke-like pattern appearing to emanate from IRAS 06382+0939
and coinciding with submillimeter dust filaments. Additionally they
found that the nearest neighbor spacing between these sources had a
peak in the distribution of approximately 20\arcsec, very close to
the Jeans length of the cloud.  They argued that the 24 $\mu$m
sources represented collapsing, fragmenting filaments.

The brightest of these cores in the submillimeter is IRAS 12 S1,
located near the position of the \citet{Cohen85} secondary peak.
IRAS 12 S1 also corresponds to the millimeter source D-MM1 in the
survey of \citet*{Peretto06}. \citet{Wolf-Chase03} suggested that
IRAS 12 S1 is an intermediate mass Class 0 protostar.    In this
paper we present {\it Spitzer} and Magellan observations that
demonstrate that S1 is, in fact, a complex of extremely young
objects.

\section{ New Observations}
\subsection{{\it Spitzer} Observations}

NGC 2264 was observed with IRAC \citep{Fazio04} on 2004 March 6. The
observations were conducted in High Dynamic Range mode, with
integration times of 0.4 s and 10.4 s per frame.  Three integrations
were taken at each dithered position, yielding a typical total
integration time of 31 s per point.  The grid was laid out in a 7
$\times$ 11 pattern with 290\arcsec \/ offsets, resulting in a total
coverage area of approximately 33\arcmin $\times$51\arcmin. The map
was centered at 06$^{h}$40$^{m}$54\fs92
+09\arcdeg37\arcmin07\farcs94 \/ with a position angle of
-1.5\arcdeg. In High Dynamic Range mode, each position has both long
and short integrations to allow the non-saturating measurement of
both bright and faint sources.  The standard Basic Calibrated Data
from the {\it Spitzer} Science Center were mosaicked using IRAF
routines, and source photometry was obtained using DAOPHOT and IDL.

The MIPS \citep{Rieke04} observations were conducted on 2004 March
16 using the scan map mode.  Fourteen scan legs of 0.75\arcdeg  \/
length and 160\arcsec  \/ offsets were taken at medium speed.  Total
integration times of 80 s per point and 40 s per point were obtained
in the 24 $\mu$m band and 70 $\mu$m band, respectively.  We also
obtained sparse coverage in the 160 $\mu$m band, but most of the
data were saturated due to the extremely high backgrounds in the
molecular cloud. The full map was centered at
6$^{h}$40$^{m}$55$^{s}$ +9\arcdeg37\arcmin08\arcsec \/ with a
position angle of 179\arcdeg. These observations were processed with
the MIPS Data Analysis Tool (DAT) \citep{Gordon05} which produces
calibrated mosaics of the mapped regions. Processing of the
resultant image products to obtain photometry was done using
standard routines in DAOPHOT and IDL.

For this paper, we utilize a magnitude scale based on the spectral
energy distribution of an A0 star.  The magnitude zero points for
the 3.6, 4.5, 5.8, 8.0, and 24 $\mu$m bands are 277.5, 179.5, 116.6,
63.1, and 7.3 Jy, respectively.

\subsection{Magellan Observations}

Near infrared imaging of the IRAS 12 S1 region was obtained with the
PANIC camera on the 6.5-m Baade telescope \citep{Martini04} at Las
Campanas Observatory. The PANIC camera uses a 2.5 $\mu$m cutoff
HAWAII-1 HgCdTe array with a scale of 0.125\arcsec \/ per pixel.
Images were taken in the $J$, $H$, and $K_s$ bands as well as in a
narrow band molecular hydrogen filter centered on the 2.12 $\mu$m
1-0 S(1) line.  The broadband images were obtained on 2004 December
30 with a total integration time of 900 s in each band. The
molecular hydrogen observations were taken on 2005 February 23 with
a total integration time of 2400 s.  Photometric calibration of
PANIC images was done using a suite of 2MASS sources detected in the
images.  The observations were obtained under excellent conditions,
with a stellar FWHM of 0.37\arcsec for the December observations and
0.44\arcsec for the February run. Processing of the PANIC images was
done with IRAF scripts developed for that instrument by S.E. Persson
and P. Martini, and source photometry was done with standard DAOPHOT
routines.  Point source photometry was obtained with fitting of an
empirical point spread function.

\section{Results}
Figure 1 is a color composite image of the IRS-2 region utilizing
both IRAC and MIPS data.  The color coding is blue = 3.6 $\mu$m,
green = 8.0 $\mu$m, and red = 24 $\mu$m.  We have chosen this
combination of bands to highlight the different kinds of physical
objects in the region.  The 3.6 $\mu$m emission is dominated by
stellar photospheres, and the majority of the sources found in this
band are also detected in the near infrared at 2MASS sensitivities.
The 8 $\mu$m emission is associated with hot dust, and for a region
as young as NGC 2264, most of the 8 $\mu$m emitters are T Tauri
stars.  There can also be a substantial extended component to the 8
$\mu$m emission that is due to the strong PAH band at this
wavelength. Finally, the 24 $\mu$m band is particularly useful for
identifying the Class I and Class 0 objects --- protostars with
significant circumstellar disks and envelopes.

The emission in the region is dominated by IRS-2.  The source
consists of an unresolved component, present at all wavelengths and
an extended component that is particularly prominent at 8 $\mu$m.
The 24 $\mu$m emission highlights the coldest and presumably
youngest objects in the region. They appear to be aligned like the
spokes of a wheel.  This overall morphology and its relationship
with the dense cloud detected at submillimeter wavelengths is
discussed in detail in the paper by \citet{Teixeira06}.

The present paper focuses on the very dense collection of sources
denoted by "core" in Figure 1.  In the IRAC observations at 3.6,
4.5, 5.8, and 8.0 $\mu$m, the sources are somewhat confused given
the angular resolution of approximately 2\arcsec \/ afforded by {\it
Spitzer} at IRAC wavelengths. We are still able to discern at least
10 distinct sources at 3.6 $\mu$m within a diameter of 20\arcsec,
which at the assumed distance of NGC 2264 corresponds to only 0.08
pc. The implied stellar density is greater than 4 x 10$^{4}$
pc$^{-3}$.

Figure 2 shows a $JHK_s$ color composite taken with the PANIC
camera. The image covers the 1.8\arcmin \/ x 1.6\arcmin \/ box
indicated on Figure 1.  Both the core cluster and a small
surrounding region are included in the image. The excellent seeing
in the PANIC observations reveals a complicated mixture of point
sources, extended emission, and obscuration bands. Figure 3 shows
the image of the same region taken in the 2.12 $\mu$m molecular
hydrogen line. We have performed a crude continuum correction by
subtracting a $K_s$ band image that has been scaled by the relative
bandwidths of the filters. While this subtraction is good enough to
identify pure molecular hydrogen features, it is inaccurate for
identifying continuum sources that also emit in $H_2$.  We find that
there are extended regions of $H_2$ emission, particularly to the
west of the core.

Photometry from the {\it Spitzer} and PANIC observations is
presented in Table 1.  The source list covers the region shown in
the PANIC image.  Given the very wide range of angular resolution in
this data set (from FWHM of 0.37 \arcsec \/ for the PANIC images to
7\arcsec \/for the 24 $\mu$m band), it is not possible in every case
to unambiguously assign the appropriate long wavelength flux to each
PANIC source. When possible, we use data from the PANIC data for
$JHK_s$ magnitudes.  Sources brighter than $J=12$, $H=12$, and
$K_s=11$ are saturated, and we substitute 2MASS photometry. Sources
detected in the PANIC observations are indicated in Figure 4 with
numbers below 100.  Sources detected only in the {\it Spitzer}
images have numbers greater than 1000.

\section{Discussion}

The simulations of \citet{Bate03} and \citet{Bate05} follow the
collapse of a Jeans unstable turbulent core.  They show that star
formation commences on time scales comparable to the global
free-fall time, and it proceeds by forming dense cores that fragment
into stars and brown dwarfs.  Dynamical interactions are important
in the development of the ensemble.  More recently,
\citet{Kurosawa04} have simulated the appearance of the
\citet{Bate03} simulations in the {\it Spitzer} bands.  We will
argue that the S1 stars represent just such a fragmenting core. For
this hypothesis to be plausible, it will be necessary to demonstrate
the extreme youth of the objects.

Figure 4 is a $K_s$ PANIC image of the S1 region with the sources
identified.  Overlaying the image are the 850 $\mu$m contours from
\citet{Wolf-Chase03}.   As shown in the figure, the "micro-cluster"
is located precisely at the peak of the IRAS-12 S1 submillimeter
source mapped by \citet{Wolf-Chase03} (also denoted Core C by
\citet{Williams02}). The entire core has a composite spectral energy
distribution (SED) that peaks in the far-infrared.
\citet{Wolf-Chase03} derived an envelope mass of 17.6 $M_\sun$
within a 28\arcsec \/diameter. Based on the ratio of submillimeter
to bolometric luminosity, they classified S1 as an intermediate-mass
Class 0 protostar. \citet{Williams02} observed the region in $HCO^+$
and detected an emission line with a strong, red-shifted absorption
feature. They interpreted this spectral signature as indicative of
infall.  Hence, both continuum and line observations suggest that
IRAS 12 S1 is undergoing active collapse. The submillimeter contours
of \citet{Wolf-Chase03} show that S1 is somewhat extended but
unresolved at both 850 and 450 \micron.  Rather than being a single
protostar, however, it is clear from the new shorter wavelength
observations that S1 is made up of a cluster of discrete sources.

Figure 5 shows the SEDs of the sources encompassed by the
submillimeter peak. Except for \#34, which appears to be a
unreddened foreground star, all the sources are steeply rising to
longer wavelengths and can be classified as Class 0 or Class I.
While all these stars appear to have significant amounts of
circumstellar matter, it is not possible to unambiguously assign the
relative contributions of the stars to the 24 $\mu$m flux since in
many cases the sources are badly confused.  For example, the flux
given to \#38 really is a composite that can include contributions
from \#41 and others.

The SEDs in S1 are markedly different from those just outside the
dense submillimeter core. Figures 6 and 7 show the SEDs for the
regions 1\arcmin \/ West and 1\arcmin \/ Northeast of S1,
respectively. There are a number of flat SEDs in those two adjacent
regions, but the majority of the detected objects have SEDs
consistent with highly reddened low-mass stars.

A second indication of the youth of the micro-cluster is the
presence of extensive molecular hydrogen flows.  In the S1 core
itself, several knots of molecular hydrogen are present, for
example, between sources \#47 and \#49 and between sources \#40 and
\#37. The most extensive molecular hydrogen line emission is away
from the core, however.  A north-south series of knots (indicated by
the arrows in Figure 3) is quite prominent in the continuum
subtracted $H_2$ image. The morphology is reminiscent of bipolar
Herbig-Haro flows. However, we find no evidence for any candidate
sources at 24 or 70 $\mu$m other than possibly \#26.  It is a
relatively weak flat-spectrum source with no significant emission at
70 $\mu$m.  It is possible that one of the Class I sources in the
micro-cluster is the source of the $H_2$ emission, although that
interpretation would be difficult to reconcile with the North-South
geometry of the knots.  That possibility will likely require proper
motion studies of the knots to prove conclusively.

A number of the sources appear to have opaque dust lanes that are
likely caused by circumstellar disks.  The best example is \#44,
which has a distinctly bipolar appearance.  Moreover, this source is
associated with molecular hydrogen emission that emanates from the
"pole" and terminates in a bright knot 7\arcsec to the northeast.

An upper limit to the age of the S1 group can be determined from
dynamical arguments.  The characteristic non-thermal velocity
dispersion of the gas in the core has been measured by
\citet{Williams02} to be 0.7 - 1.0 km s$^{-1}$. \citet{Williams02}
also identified five other submillimeter cores in the main S1
complex.  The measured core to core velocity dispersion is 0.9 km
s$^{-1}$. If we adopt a 0.7 km s$^{-1}$ 1-dimensional dispersion for
the stars, the dispersal time in the S1 micro-cluster is only 40000
years for a 10\arcsec radius. The 0.7 km s$^{-1}$ velocity
dispersion may, in fact, be an underestimate if we consider velocity
dispersions found in other star formation regions. For example,
\citet{Jones88} measured a 1-dimensional dispersion of 2 km s$^{-1}$
for the Orion nebula stars.  The simulations of turbulent core
collapse exhibit many examples of dynamical interactions between the
forming stars. In particular, the lower mass members often attain a
high enough velocity to be ejected from the core region.
\citet{Bate05} found 1-dimensional velocity dispersions between 1
and 2.5 km s$^{-1}$, depending on initial conditions. Hence, our use
of 0.7 km s$^{-1}$ for the velocity dispersion is likely to be
conservative.

What can be said about the masses of the members of the
micro-cluster?  Unfortunately, the wide range of parameters afforded
by extinction (both disk and envelope), inclination, and accretion
rate makes fitting of individual sources to evolutionary models
problematic.  Limits can be established, however, if we consider the
ensemble properties of the micro-cluster.  \citet{Wolf-Chase03}
derive an envelope mass of 17.6 $M_{\sun}$ and a bolometric
luminosity of 107.5 $L_{\sun}$, leading to their association of the
source with an intermediate mass protostar.  The {\it Spitzer}
observations show, however, that the luminosity is actually shared
by perhaps a dozen lower-mass objects.  For extremely young objects,
the far-infrared and submillimeter luminosity is expected to be a
valid measure of the bolometric luminosity since they will still be
surrounded by envelopes that intercept the bulk of the short
wavelength emission.  Applying the pre-main sequence tracks of
\citet{Siess00}, a star with a luminosity of 10 $L_{\sun}$
corresponds to a mass of only 0.9 $M_{\sun}$ at an age of 10$^{5}$
years.  Thus, we may be seeing the fragmentation of a molecular core
into solar or lower mass stars.

\section{Summary}

As part of a larger program surveying young clusters, we have mapped
the star forming region NGC 2264 with the MIPS and IRAC instruments
on {\it Spitzer}.  We have found a remarkable collection of
extremely young objects in the IRS-2 region \citep{Teixeira06}. One
of these objects, associated with the submillimeter source IRAS-12
S1 proves to be a dense collection of protostellar objects in its
own right. Based on the steeply rising SEDs of the objects, the
proximity of molecular hydrogen outflows, and a dynamical age of
less than 40000 years, we argue that this "micro cluster" may
represent a fragmenting, collapsing core, very much as simulated by
\citet{Bate03}.

\acknowledgements This work is based on observations made with the
{\it Spitzer} Space Telescope, which is operated by the Jet
Propulsion Laboratory, California Institute of Technology under NASA
contract 1407. Support for this work was provided by NASA through
Contract Number 960785 issued by JPL/Caltech.  This work made use of
Simbad and 2MASS databases as well as IRAF, ds9, and the IDL
Astronomy Library. P. Teixeira acknowledges support from the
scholarship SFRH/BD/13984/2003 awarded by the Fundacao para a
Ci\^encia Tecnologia (Portugal).  This paper includes data gathered
with the 6.5 meter Magellan Telescopes located at Las Campanas
Observatory, Chile.

{\it Facilities:} \facility{Spitzer ()}, \facility{Magellan:Baade
()}
%


\begin{deluxetable}{cccccccccccc}
\rotate \tabletypesize{\footnotesize} \tablewidth{0pt}
\tablecaption{Sources in the IRAS 12 S1 Core\label{Tab1}}
\footnotesize \tablehead{ \colhead{STAR
ID}&\colhead{RA(J2000)}&\colhead{DEC(J2000)}&\colhead{2MASS
NAME}&\colhead{$J$}
&\colhead{$H$}&\colhead{$K$}&\colhead{[3.6]}&\colhead{[4.5]}&\colhead{[5.8]}&\colhead{[8.0]}&\colhead{[24]}}
\startdata
2&06$^{h}$41$^{m}$01.32$^{s}$&$+$09\arcdeg34\arcmin52.7\arcsec&06410133+0934526&12.91&12.83&12.46&11.94&11.71&11.40&10.44&6.49\\
3&06$^{h}$41$^{m}$01.37$^{s}$&$+$09\arcdeg34\arcmin07.7\arcsec&&13.20&12.96&12.34&. . . &. . . &. . . &. . . &. . . \\
4&06$^{h}$41$^{m}$01.42$^{s}$&$+$09\arcdeg34\arcmin08.0\arcsec&06410141+0934081&12.35&11.71&11.61&11.40&11.40&11.28&10.97&9.03\\
6&06$^{h}$41$^{m}$01.56$^{s}$&$+$09\arcdeg34\arcmin32.9\arcsec&06410156+0934329&14.40&12.64&11.38&10.05&9.69&9.05&8.24&. . . \\
7&06$^{h}$41$^{m}$01.75$^{s}$&$+$09\arcdeg34\arcmin40.4\arcsec&&20.25&17.26&15.50&13.52&12.78&12.12&. . . &. . . \\
8&06$^{h}$41$^{m}$01.78$^{s}$&$+$09\arcdeg33\arcmin34.6\arcsec&&&&18.69&16.02&15.85&. . . &. . . &. . . \\
9&06$^{h}$41$^{m}$01.82$^{s}$&$+$09\arcdeg34\arcmin34.0\arcsec&06410182+0934342&17.08&13.80&11.67&9.31&8.61&7.76&6.99&2.93\tablenotemark{a}\\
10&06$^{h}$41$^{m}$01.97$^{s}$&$+$09\arcdeg34\arcmin30.7\arcsec&&&16.49&14.61&12.37&11.19&. . . &. . . &. . . \\
1332&06$^{h}$41$^{m}$02.18$^{s}$&$+$09\arcdeg34\arcmin14.5\arcsec&&&&&15.95&14.23&13.56&. . . &. . . \\
1334&06$^{h}$41$^{m}$02.21$^{s}$&$+$09\arcdeg34\arcmin30.0\arcsec&&&&&. . . &11.29&. . . &. . . &. . . \\
11&06$^{h}$41$^{m}$02.52$^{s}$&$+$09\arcdeg34\arcmin55.9\arcsec&06410253+0934557&12.58&11.82&11.62&11.61&11.67&11.61&11.42&. . . \\
1349&06$^{h}$41$^{m}$02.54$^{s}$&$+$09\arcdeg35\arcmin02.0\arcsec&&&&&16.33&. . . &. . . &. . . &. . . \\
1352&06$^{h}$41$^{m}$02.57$^{s}$&$+$09\arcdeg34\arcmin42.6\arcsec&&&&&. . . &. . . &14.55&. . . &. . . \\
13&06$^{h}$41$^{m}$02.59$^{s}$&$+$09\arcdeg34\arcmin18.8\arcsec&06410258+0934190&12.61&12.11&11.89&11.91&11.99&11.98&12.03&. . . \\
1374&06$^{h}$41$^{m}$02.81$^{s}$&$+$09\arcdeg34\arcmin08.0\arcsec&&&&&15.76&14.37&13.57&. . . &. . . \\
15&06$^{h}$41$^{m}$03.07$^{s}$&$+$09\arcdeg33\arcmin27.4\arcsec&&16.96&16.14&15.78&15.57&15.68&. . . &. . . &. . . \\
1388&06$^{h}$41$^{m}$03.07$^{s}$&$+$09\arcdeg34\arcmin03.0\arcsec&&&&&. . . &14.80&13.55&. . . &. . . \\
1392&06$^{h}$41$^{m}$03.19$^{s}$&$+$09\arcdeg32\arcmin55.0\arcsec&06410319+0932550&14.44&13.76&13.36&12.76&12.59&12.33&11.69&. . . \\
17&06$^{h}$41$^{m}$03.60$^{s}$&$+$09\arcdeg34\arcmin53.0\arcsec&&&18.03&16.14&14.23&13.59&13.15&. . . &. . . \\
16&06$^{h}$41$^{m}$03.60$^{s}$&$+$09\arcdeg34\arcmin04.4\arcsec&&&&18.11&16.37&15.90&. . . &. . . &. . . \\
18&06$^{h}$41$^{m}$03.79$^{s}$&$+$09\arcdeg34\arcmin53.8\arcsec&&16.74&15.87&15.54&. . . &. . . &. . . &. . . &. . . \\
1424&06$^{h}$41$^{m}$03.82$^{s}$&$+$09\arcdeg33\arcmin23.0\arcsec&&&&&15.59&14.18&13.50&. . . &. . . \\
1425&06$^{h}$41$^{m}$03.86$^{s}$&$+$09\arcdeg35\arcmin03.8\arcsec&&&&&. . . &13.27&. . . &. . . &. . . \\
1427&06$^{h}$41$^{m}$03.89$^{s}$&$+$09\arcdeg34\arcmin32.9\arcsec&&&&&16.64&15.42&13.85&. . . &. . . \\
1432&06$^{h}$41$^{m}$03.98$^{s}$&$+$09\arcdeg35\arcmin03.1\arcsec&&&&&. . . &. . . &7.57&. . . &. . . \\
1433&06$^{h}$41$^{m}$03.98$^{s}$&$+$09\arcdeg33\arcmin18.0\arcsec&&&&&. . . &15.47&. . . &. . . &. . . \\
19&06$^{h}$41$^{m}$04.06$^{s}$&$+$09\arcdeg33\arcmin48.6\arcsec&06410405+0933486&15.16&13.58&12.90&12.47&12.42&12.32&12.24&. . . \\
1438&06$^{h}$41$^{m}$04.13$^{s}$&$+$09\arcdeg33\arcmin01.4\arcsec&06410411+0933018&10.17&10.22&10.22&10.28&10.32&10.38&10.19&. . . \\
20&06$^{h}$41$^{m}$04.15$^{s}$&$+$09\arcdeg34\arcmin57.0\arcsec&06410417+0934572&14.83&14.05&13.68&. . . &. . . &. . . &. . . &. . . \\
21&06$^{h}$41$^{m}$04.20$^{s}$&$+$09\arcdeg33\arcmin37.1\arcsec&&&&16.67&15.29&13.68&13.26&. . . &. . . \\
22&06$^{h}$41$^{m}$04.20$^{s}$&$+$09\arcdeg33\arcmin23.8\arcsec&&12.58&&17.02&14.89&13.42&12.46&11.20&. . . \\
1444&06$^{h}$41$^{m}$04.22$^{s}$&$+$09\arcdeg33\arcmin32.0\arcsec&&&&&15.50&14.18&13.16&11.87&. . . \\
23&06$^{h}$41$^{m}$04.25$^{s}$&$+$09\arcdeg34\arcmin59.5\arcsec&&&15.85&12.44&9.48&8.57&7.57&7.04&3.07\\
1445&06$^{h}$41$^{m}$04.25$^{s}$&$+$09\arcdeg34\arcmin52.0\arcsec&&&&&. . . &12.30&. . . &. . . &. . . \\
1448&06$^{h}$41$^{m}$04.25$^{s}$&$+$09\arcdeg34\arcmin45.5\arcsec&&&&&. . . &14.37&12.76&. . . &. . . \\
24&06$^{h}$41$^{m}$04.30$^{s}$&$+$09\arcdeg34\arcmin00.8\arcsec&&&&17.14&15.40&14.58&13.63&. . . &. . . \\
1460&06$^{h}$41$^{m}$04.37$^{s}$&$+$09\arcdeg32\arcmin57.1\arcsec&06410436+0932576&12.75&15.09&12.68&14.30&14.14&14.23&. . . &. . . \\
1459&06$^{h}$41$^{m}$04.37$^{s}$&$+$09\arcdeg33\arcmin59.4\arcsec&&&&&15.57&. . . &. . . &. . . &. . . \\
26&06$^{h}$41$^{m}$04.46$^{s}$&$+$09\arcdeg33\arcmin43.6\arcsec&&&18.91&16.27&14.29&13.55&12.75&11.65&7.22\\
1465&06$^{h}$41$^{m}$04.46$^{s}$&$+$09\arcdeg34\arcmin50.5\arcsec&&&&&. . . &14.83&. . . &. . . &. . . \\
1470&06$^{h}$41$^{m}$04.58$^{s}$&$+$09\arcdeg34\arcmin47.6\arcsec&&&&&15.41&. . . &. . . &. . . &. . . \\
1473&06$^{h}$41$^{m}$04.63$^{s}$&$+$09\arcdeg33\arcmin54.4\arcsec&&&&&. . . &15.24&. . . &. . . &. . . \\
1474&06$^{h}$41$^{m}$04.63$^{s}$&$+$09\arcdeg34\arcmin58.1\arcsec&&&&&. . . &. . . &10.44&. . . &. . . \\
1476&06$^{h}$41$^{m}$04.70$^{s}$&$+$09\arcdeg33\arcmin48.6\arcsec&&&&&. . . &16.29&. . . &. . . &. . . \\
1478&06$^{h}$41$^{m}$04.70$^{s}$&$+$09\arcdeg34\arcmin54.8\arcsec&&&&&. . . &13.31&. . . &. . . &. . . \\
1481&06$^{h}$41$^{m}$04.78$^{s}$&$+$09\arcdeg34\arcmin43.3\arcsec&&&&&15.93&14.10&. . . &. . . &. . . \\
1485&06$^{h}$41$^{m}$04.85$^{s}$&$+$09\arcdeg35\arcmin03.5\arcsec&&&&&. . . &14.61&. . . &. . . &. . . \\
1492&06$^{h}$41$^{m}$05.06$^{s}$&$+$09\arcdeg33\arcmin00.0\arcsec&06410505+0933002&14.13&13.37&13.16&12.88&12.98&12.86&12.46&. . . \\
30&06$^{h}$41$^{m}$05.16$^{s}$&$+$09\arcdeg34\arcmin10.2\arcsec&&&17.42&15.16&12.05&10.83&9.94&9.07&. . . \\
33&06$^{h}$41$^{m}$05.35$^{s}$&$+$09\arcdeg33\arcmin13.3\arcsec&06410536+0933134&12.58&11.85&11.64&11.41&11.55&11.38&11.38&. . . \\
32&06$^{h}$41$^{m}$05.35$^{s}$&$+$09\arcdeg34\arcmin13.4\arcsec&06410535+0934133&19.42&16.09&13.95&11.69&10.91&9.71&. . . &. . . \\
1512&06$^{h}$41$^{m}$05.40$^{s}$&$+$09\arcdeg33\arcmin23.4\arcsec&&&&&16.52&15.86&. . . &. . . &. . . \\
34&06$^{h}$41$^{m}$05.42$^{s}$&$+$09\arcdeg34\arcmin09.5\arcsec&06410542+0934095&14.66&13.77&13.48&13.25&. . . &. . . &. . . &. . . \\
35&06$^{h}$41$^{m}$05.50$^{s}$&$+$09\arcdeg34\arcmin11.6\arcsec&&&&15.63&11.85&. . . &9.68&8.92&. . . \\
1515&06$^{h}$41$^{m}$05.50$^{s}$&$+$09\arcdeg34\arcmin36.8\arcsec&&&&&. . . &15.40&. . . &. . . &. . . \\
36&06$^{h}$41$^{m}$05.52$^{s}$&$+$09\arcdeg35\arcmin01.3\arcsec&&&16.56&15.00&13.88&13.70&13.27&12.60&. . . \\
38&06$^{h}$41$^{m}$05.57$^{s}$&$+$09\arcdeg34\arcmin08.0\arcsec&&&&16.02&11.73&10.43&9.40&8.74&2.21\tablenotemark{b}\\
39&06$^{h}$41$^{m}$05.62$^{s}$&$+$09\arcdeg33\arcmin55.1\arcsec&06410562+0933549&16.72&14.63&13.33&11.82&11.23&10.60&9.77&. . . \\
41&06$^{h}$41$^{m}$05.74$^{s}$&$+$09\arcdeg34\arcmin06.2\arcsec&&&19.01&15.07&11.73&10.47&9.46&8.71&. . . \\
42&06$^{h}$41$^{m}$05.76$^{s}$&$+$09\arcdeg33\arcmin47.9\arcsec&&17.88&16.82&16.12&14.87&14.06&13.05&. . . &. . . \\
1530&06$^{h}$41$^{m}$05.76$^{s}$&$+$09\arcdeg33\arcmin00.4\arcsec&&&&&16.21&15.33&14.48&. . . &. . . \\
1542&06$^{h}$41$^{m}$05.86$^{s}$&$+$09\arcdeg34\arcmin52.7\arcsec&&&&&. . . &14.23&. . . &. . . &. . . \\
43&06$^{h}$41$^{m}$05.88$^{s}$&$+$09\arcdeg34\arcmin45.8\arcsec&&16.75&13.66&12.15&11.18&11.00&10.87&10.65&. . . \\
44&06$^{h}$41$^{m}$05.93$^{s}$&$+$09\arcdeg34\arcmin11.3\arcsec&06410598+0934115&&17.35&15.30&12.90&11.75&10.73&9.44&3.8\tablenotemark{c}\\
45&06$^{h}$41$^{m}$05.93$^{s}$&$+$09\arcdeg34\arcmin04.8\arcsec&&&&19.97&. . . &. . . &10.29&8.68&. . . \\
1547&06$^{h}$41$^{m}$05.93$^{s}$&$+$09\arcdeg35\arcmin03.1\arcsec&&&&&13.78&13.49&13.37&11.89&. . . \\
1549&06$^{h}$41$^{m}$05.95$^{s}$&$+$09\arcdeg34\arcmin39.7\arcsec&&&&&. . . &13.36&. . . &. . . &. . . \\
46&06$^{h}$41$^{m}$06.05$^{s}$&$+$09\arcdeg34\arcmin46.2\arcsec&06410604+0934461&16.40&13.58&12.25&11.39&11.18&10.99&10.60&. . . \\
47&06$^{h}$41$^{m}$06.17$^{s}$&$+$09\arcdeg34\arcmin08.8\arcsec&&&&15.86&12.51&10.58&9.83&9.19&. . . \\
1561&06$^{h}$41$^{m}$06.24$^{s}$&$+$09\arcdeg33\arcmin08.6\arcsec&06410623+0933087&14.11&13.35&13.05&12.55&12.44&12.39&12.22&. . . \\
48&06$^{h}$41$^{m}$06.29$^{s}$&$+$09\arcdeg33\arcmin50.0\arcsec&06410629+0933496&&19.04&16.73&11.44&10.06&8.90&7.91&3.33\\
1568&06$^{h}$41$^{m}$06.34$^{s}$&$+$09\arcdeg34\arcmin15.2\arcsec&&&&&. . . &12.97&11.08&. . . &. . . \\
49&06$^{h}$41$^{m}$06.38$^{s}$&$+$09\arcdeg34\arcmin10.2\arcsec&06410642+0934099&&16.44&14.48&12.59&. . . &. . . &. . . &. . . \\
1576&06$^{h}$41$^{m}$06.41$^{s}$&$+$09\arcdeg35\arcmin06.0\arcsec&06410640+0935061&16.42&15.95&15.22&15.31&14.87&. . . &. . . &. . . \\
1579&06$^{h}$41$^{m}$06.43$^{s}$&$+$09\arcdeg33\arcmin43.2\arcsec&&&&&14.87&13.52&12.81&. . . &. . . \\
1588&06$^{h}$41$^{m}$06.60$^{s}$&$+$09\arcdeg35\arcmin00.2\arcsec&&&&&. . . &. . . &. . . &12.15&. . . \\
1591&06$^{h}$41$^{m}$06.60$^{s}$&$+$09\arcdeg35\arcmin02.0\arcsec&&&&&. . . &. . . &13.56&. . . &. . . \\
51&06$^{h}$41$^{m}$06.62$^{s}$&$+$09\arcdeg34\arcmin21.0\arcsec&&19.16&16.89&15.55&14.15&13.66&12.75&11.59&. . . \\
52&06$^{h}$41$^{m}$06.65$^{s}$&$+$09\arcdeg33\arcmin57.6\arcsec&06410665+0933576&&16.99&13.87&11.57&10.73&9.98&9.12&5.1\\
1597&06$^{h}$41$^{m}$06.65$^{s}$&$+$09\arcdeg34\arcmin26.4\arcsec&&&&&16.81&15.02&. . . &. . . &. . . \\
53&06$^{h}$41$^{m}$06.70$^{s}$&$+$09\arcdeg33\arcmin29.9\arcsec&&&&17.06&13.70&12.61&. . . &. . . &. . . \\
1600&06$^{h}$41$^{m}$06.72$^{s}$&$+$09\arcdeg34\arcmin37.9\arcsec&&&&&. . . &11.88&. . . &. . . &. . . \\
1601&06$^{h}$41$^{m}$06.72$^{s}$&$+$09\arcdeg34\arcmin31.8\arcsec&&&&&14.42&14.42&. . . &. . . &. . . \\
54\tablenotemark{d}&06$^{h}$41$^{m}$06.74$^{s}$&$+$09\arcdeg34\arcmin45.8\arcsec&06410673+0934459&11.76&10.28&9.23&8.39&7.80&6.69&5.72&1.82\\
1603&06$^{h}$41$^{m}$06.74$^{s}$&$+$09\arcdeg34\arcmin59.2\arcsec&&&&&14.29&. . . &. . . &. . . &. . . \\
1604&06$^{h}$41$^{m}$06.74$^{s}$&$+$09\arcdeg34\arcmin53.8\arcsec&&&&&. . . &11.18&. . . &. . . &. . . \\
1607&06$^{h}$41$^{m}$06.77$^{s}$&$+$09\arcdeg33\arcmin34.6\arcsec&&&&&12.06&9.45&8.70&8.47&3.54\\
1610&06$^{h}$41$^{m}$06.82$^{s}$&$+$09\arcdeg35\arcmin02.4\arcsec&&&&&. . . &. . . &13.61&. . . &. . . \\
1622&06$^{h}$41$^{m}$07.10$^{s}$&$+$09\arcdeg34\arcmin42.2\arcsec&&&&&8.37&. . . &8.42&. . . &. . . \\
1623&06$^{h}$41$^{m}$07.13$^{s}$&$+$09\arcdeg34\arcmin44.8\arcsec&&&&&. . . &. . . &8.35&. . . &. . . \\
55&06$^{h}$41$^{m}$07.20$^{s}$&$+$09\arcdeg34\arcmin24.2\arcsec&06410719+0934242&&17.93&14.81&13.16&12.67&12.34&12.13&. . . \\
1630&06$^{h}$41$^{m}$07.22$^{s}$&$+$09\arcdeg34\arcmin41.2\arcsec&&&&&. . . &12.35&. . . &. . . &. . . \\
1637&06$^{h}$41$^{m}$07.34$^{s}$&$+$09\arcdeg34\arcmin40.4\arcsec&&&&&. . . &. . . &12.10&. . . &. . . \\
56&06$^{h}$41$^{m}$07.39$^{s}$&$+$09\arcdeg34\arcmin54.8\arcsec&06410740+0934549&18.52&14.48&12.09&10.25&9.64&9.27&8.68&. . . \\
57&06$^{h}$41$^{m}$07.44$^{s}$&$+$09\arcdeg34\arcmin33.6\arcsec&06410744+0934335&15.20&14.24&13.98&13.80&13.66&13.42&. . . &. . . \\
58&06$^{h}$41$^{m}$07.49$^{s}$&$+$09\arcdeg33\arcmin37.1\arcsec&06410748+0933369&14.43&13.42&13.07&12.69&12.46&12.09&11.28&. . . \\
1656&06$^{h}$41$^{m}$07.68$^{s}$&$+$09\arcdeg34\arcmin19.2\arcsec&&&&&13.57&11.19&10.03&9.16&4.29\\
1665&06$^{h}$41$^{m}$07.92$^{s}$&$+$09\arcdeg33\arcmin33.8\arcsec&&&&&. . . &14.70&. . . &. . . &. . . \\
1666&06$^{h}$41$^{m}$07.92$^{s}$&$+$09\arcdeg34\arcmin27.8\arcsec&&&&&. . . &15.32&. . . &. . . &. . . \\
1667&06$^{h}$41$^{m}$07.92$^{s}$&$+$09\arcdeg34\arcmin33.6\arcsec&&&&&. . . &16.04&. . . &. . . &. . . \\
60&06$^{h}$41$^{m}$07.97$^{s}$&$+$09\arcdeg34\arcmin46.9\arcsec&06410798+0934469&13.97&12.53&12.06&11.05&10.57&10.20&9.41&. . . \\
1669&06$^{h}$41$^{m}$07.97$^{s}$&$+$09\arcdeg34\arcmin16.3\arcsec&&&&&. . . &13.53&. . . &. . . &. . . \\
61&06$^{h}$41$^{m}$08.06$^{s}$&$+$09\arcdeg33\arcmin13.3\arcsec&&19.05&17.91&17.87&17.42&. . . &. . . &. . . &. . . \\
1675&06$^{h}$41$^{m}$08.11$^{s}$&$+$09\arcdeg34\arcmin30.0\arcsec&&&&&16.44&15.12&. . . &. . . &. . . \\
1677&06$^{h}$41$^{m}$08.14$^{s}$&$+$09\arcdeg34\arcmin16.0\arcsec&&&&&13.26&. . . &. . . &. . . &. . . \\
62&06$^{h}$41$^{m}$08.21$^{s}$&$+$09\arcdeg34\arcmin09.5\arcsec&06410821+0934094&13.13&11.90&11.41&10.95&10.65&10.26&9.33&5.25\tablenotemark{d}\\
63&06$^{h}$41$^{m}$08.28$^{s}$&$+$09\arcdeg34\arcmin06.6\arcsec&&&&17.03&. . . &. . . &. . . &11.8&. . . \\
1683&06$^{h}$41$^{m}$08.33$^{s}$&$+$09\arcdeg34\arcmin25.7\arcsec&&&&&. . . &15.08&. . . &. . . &. . . \\
64&06$^{h}$41$^{m}$08.54$^{s}$&$+$09\arcdeg34\arcmin13.1\arcsec&06410854+0934132&16.32&13.23&11.59&10.95&10.74&10.45&10.06&. . . \\
1690&06$^{h}$41$^{m}$08.57$^{s}$&$+$09\arcdeg34\arcmin08.0\arcsec&&&&&. . . &12.53&. . . &. . . &. . . \\
\enddata
\tablenotetext{a}{The 24 $\mu$m flux is confused and contains
contributions from source 10.} \tablenotetext{b}{The 24 $\mu$m flux
is confused and contains contributions from sources 30, 32, 34, 35,
37, 40, and 41.} \tablenotetext{c}{The 24 $\mu$m flux is confused
and contains contributions form sources 47 and 49.}
\tablenotetext{d}{$=$ V 608 Mon} \tablenotetext{e}{The 24$\mu$m flux
is confused and contains contributions from the nearby source 62,
which has the more steeply rising SED.}
\end{deluxetable}

\begin{figure}
\plotone{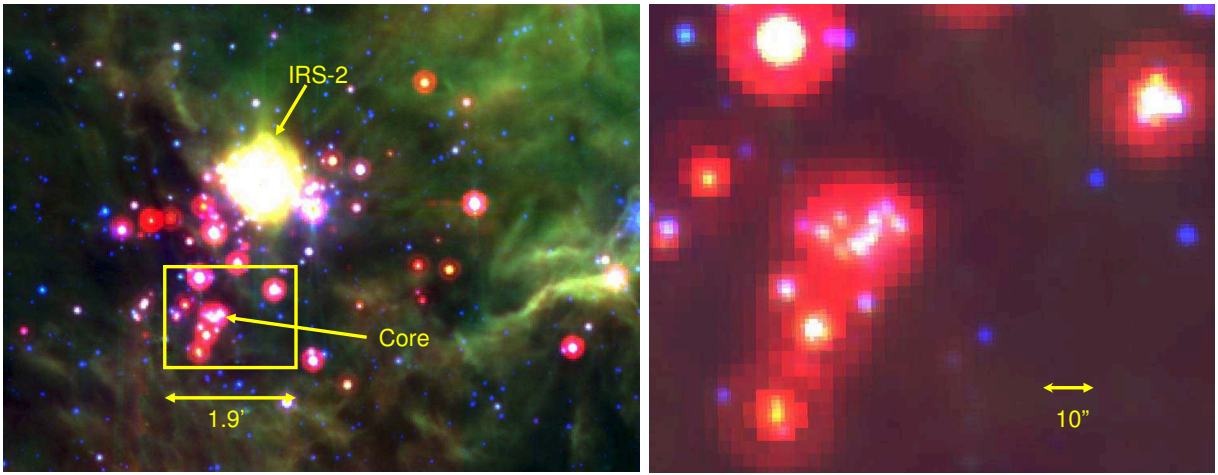} \caption{ {\it (Left)} Composite image of
{\it Spitzer} observations of the NGC 2264 IRS-2 region using data
from the 3.6 $\mu$m,  8.0 $\mu$m,  24 $\mu$m bands. The image covers
approximately 8$\farcm$1 x 5$\farcm$8. The coverage of the PANIC
image is indicated by the box. The field center is
(J2000)06$^{h}$40$^{m}$56$^{s}$ +09\arcdeg35\arcmin20\arcsec.  North
is to the top and East is to the left. {\it (Right)} Magnified
region surrounding the micro cluster. (For the electronic version,
the color coding is blue, green, and red for the three {\it Spitzer}
bands, respectively.)}
\end{figure}

\begin{figure}
\plotone{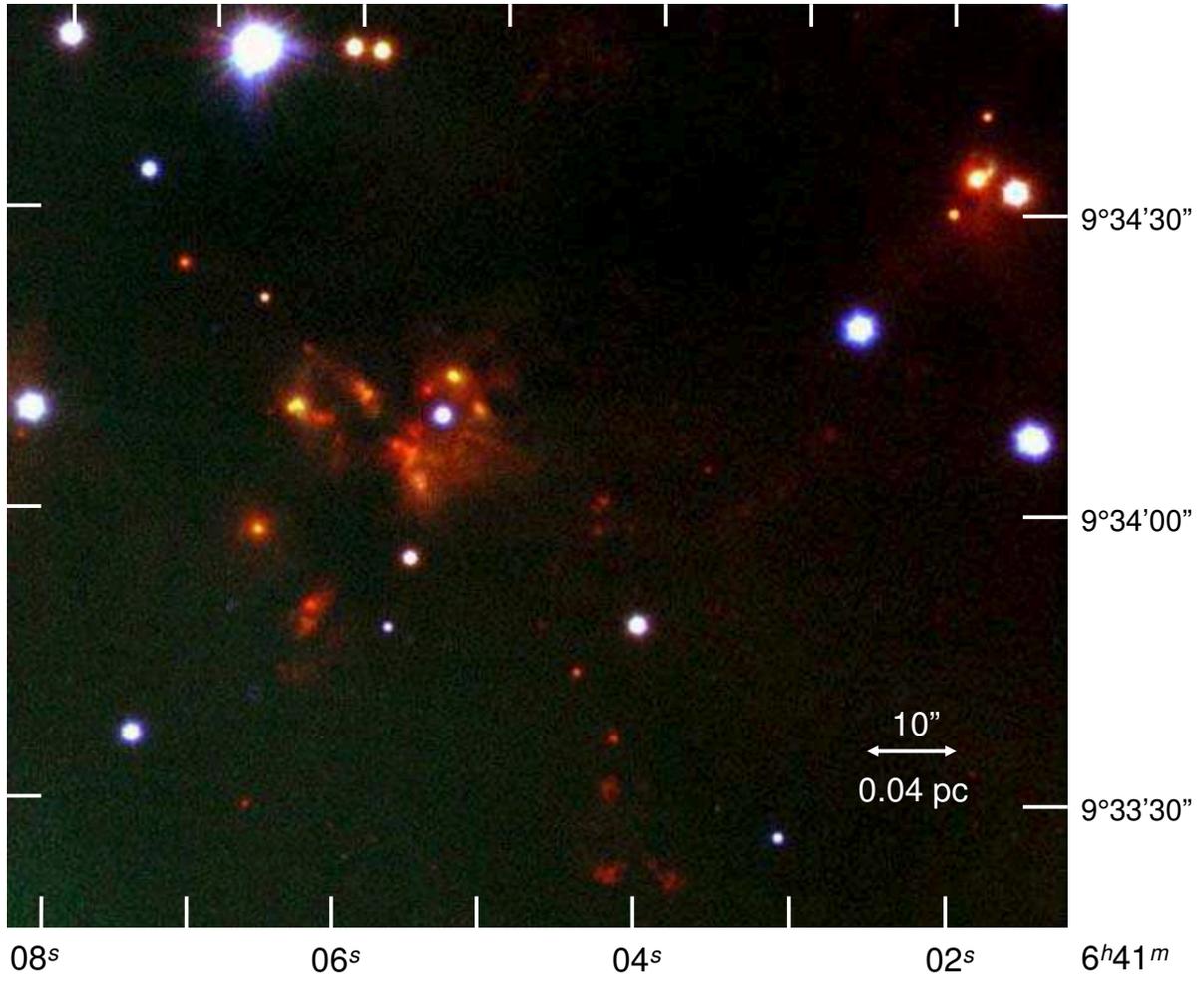} \caption{Composite image of PANIC data for the
IRAS-12 S1 Core region. (For the electronic version, the color code
is blue = $J$, green = $H$, and red = $K$.)}
\end{figure}

\begin{figure}
\plotone{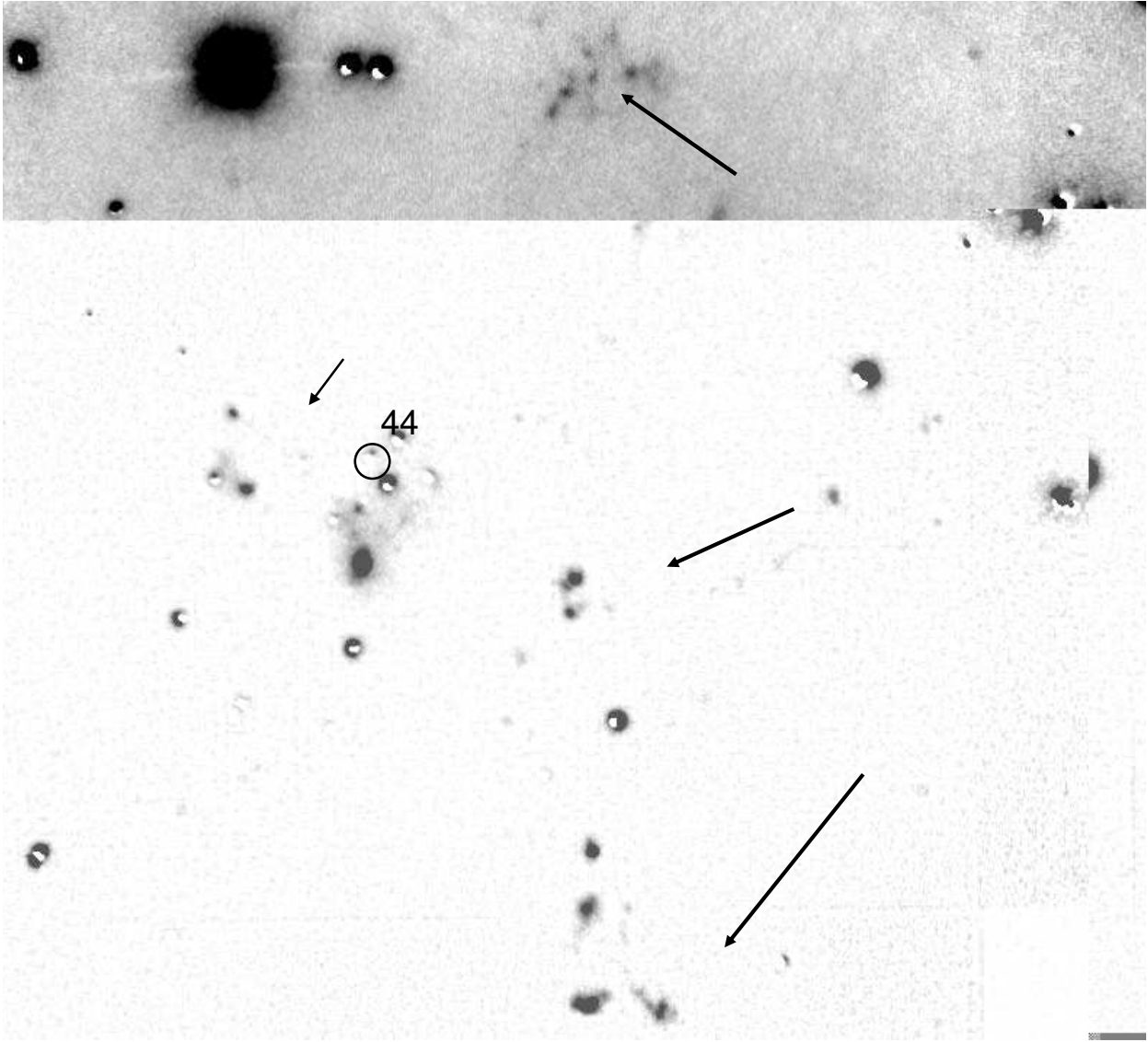} \caption{Continuum-subtracted narrow band image of
the IRAS-12 S1 Core in molecular hydrogen $S(1)$ 2.12 $\mu$m. A
prominent series of $H_2$ knots is indicated by the arrows.  Source
44 (see text) is indicated, as well as an apparently associated
molecular hydrogen knot.}
\end{figure}

\begin{figure}
\plotone{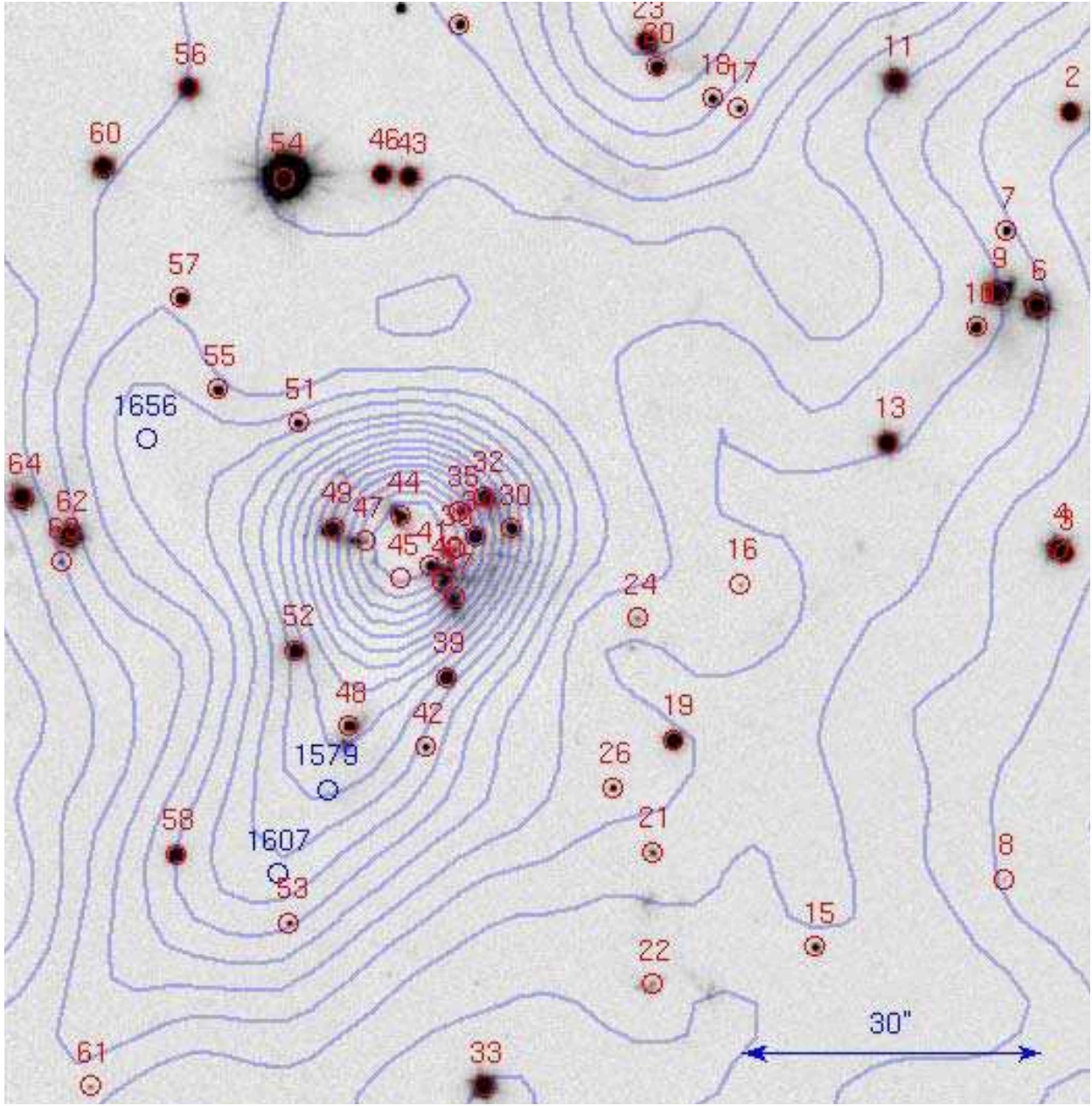} \caption{PANIC $K$-band image of the Core region.
Source numbers are indicated, with sources detected only in the {\it
Spitzer} images having numbers greater than 1000. The contours are
from 850 $\mu$m continuum SCUBA archival data \citep{Wolf-Chase03}.
 The contour intervals are 0.1 Jy/beam.}
\end{figure}

\begin{figure}
\plotone{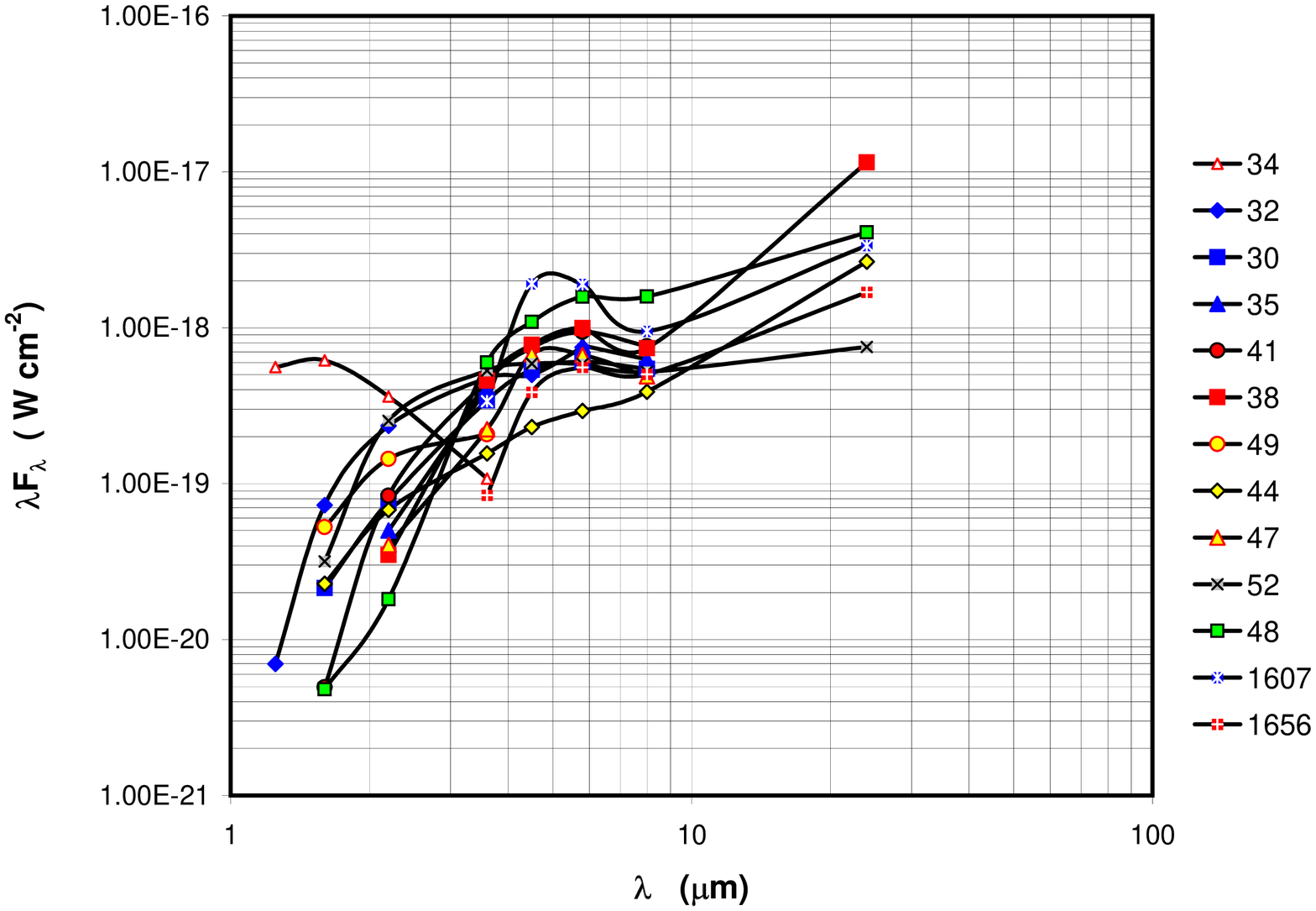} \caption{Spectral Energy Distributions for
sources in the S1 Core.  The legend numbers refer to the STAR ID
given in Table 1.}
\end{figure}

\begin{figure}
\plotone{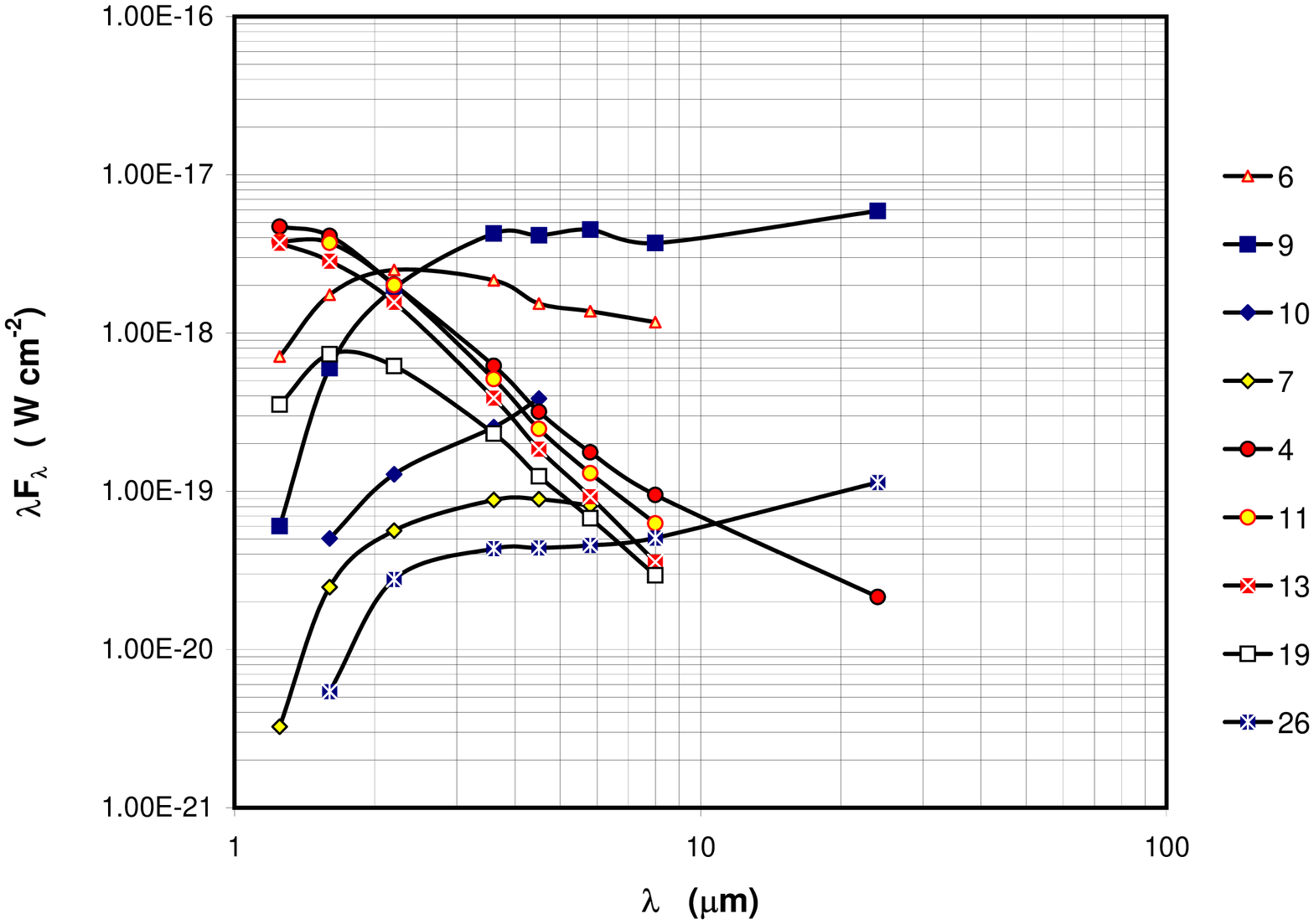} \caption{Spectral Energy Distributions for
sources west of the S1 Core.The legend numbers refer to the STAR ID
given in Table 1.}
\end{figure}

\begin{figure}
\plotone{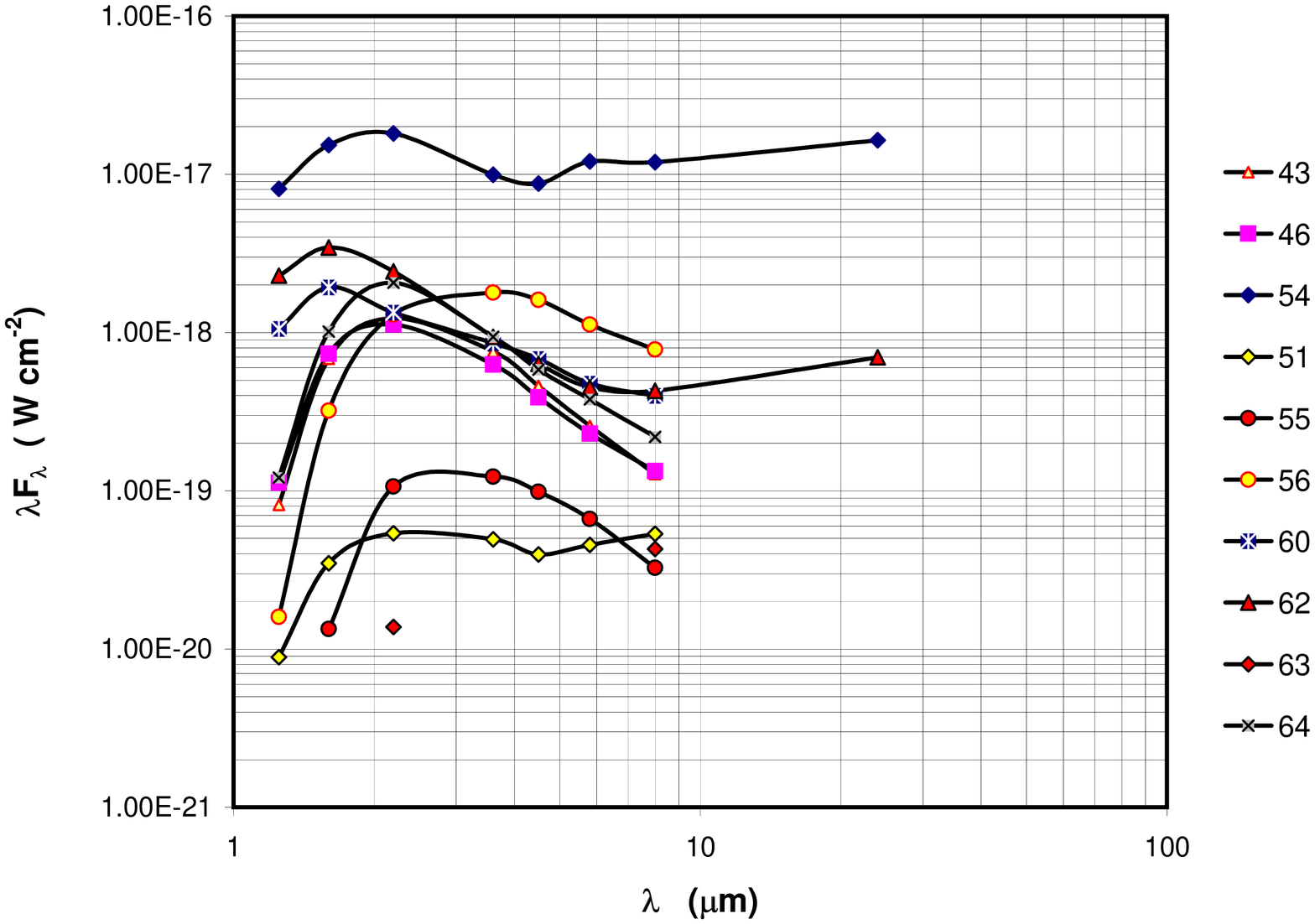} \caption{Spectral Energy Distributions for
sources northeast of the S1 Core.The legend numbers refer to the
STAR ID given in Table 1.}
\end{figure}

\end{document}